\documentclass[11pt]{article}
\usepackage{bm}
\usepackage{fancyhdr}
\usepackage{listings}
\usepackage{amsfonts}
\usepackage{amsmath}
\usepackage{mathtools}
\usepackage{geometry}
\usepackage{amssymb}
\usepackage{centernot}
\usepackage{url}
\usepackage[utf8]{inputenc}
\usepackage[english]{babel}
\usepackage{setspace}
\usepackage{listings}
\usepackage{xcolor}
\usepackage{subcaption}
\usepackage{float}
\usepackage{booktabs}
\usepackage{amsthm}

\DeclarePairedDelimiter\abs{\lvert}{\rvert}
\DeclarePairedDelimiter\norm{\lVert}{\rVert}

\makeatletter
\let\oldabs\abs
\def\abs{\@ifstar{\oldabs}{\oldabs*}}
\let\oldnorm\norm
\def\norm{\@ifstar{\oldnorm}{\oldnorm*}}
\makeatother

\makeatletter
\@tfor\next:=abcdefghijklmnopqrstuvwxyzABCDEFGHIJKLMNOPQRSTUVWXYZ\do{%
  \def\command@factory#1{%
    \expandafter\def\csname B#1\endcsname{\mathbf{#1}}
  }
 \expandafter\command@factory\next
}

\@tfor\next:=ABCDEFGHIJKLMNOPQRSTUVWXYZ\do{%
  \def\command@factory#1{%
    \expandafter\def\csname mc#1\endcsname{\mathcal{#1}}
  }
 \expandafter\command@factory\next
}

\@tfor\next:=ABCDEFGHIJKLMNOPQRSTUVWXYZ\do{%
  \def\command@factory#1{%
    \expandafter\def\csname mb#1\endcsname{\mathbb{#1}}
  }
 \expandafter\command@factory\next
}

\@tfor\next:=ABCDEFGHIJKLMNOPQRSTUVWXYZ\do{%
  \def\command@factory#1{%
    \expandafter\def\csname ms#1\endcsname{\mathsf{#1}}
  }
 \expandafter\command@factory\next
}
\makeatother

\definecolor{codegreen}{rgb}{0,0.6,0}
\definecolor{codegray}{rgb}{0.5,0.5,0.5}
\definecolor{codepurple}{rgb}{0.58,0,0.82}
\definecolor{backcolour}{rgb}{0.95,0.95,0.92}

\lstdefinestyle{mystyle}{
  backgroundcolor=\color{backcolour},   commentstyle=\color{codegreen},
  keywordstyle=\color{magenta},
  numberstyle=\tiny\color{codegray},
  stringstyle=\color{codepurple},
  basicstyle=\ttfamily\footnotesize,
  breakatwhitespace=false,         
  breaklines=true,                 
  captionpos=b,                    
  keepspaces=true,                 
  numbers=left,                    
  numbersep=5pt,                  
  showspaces=false,                
  showstringspaces=false,
  showtabs=false,                  
  tabsize=2
}

\usepackage{graphicx}
\usepackage{sidecap}
\usepackage{setspace} 
\usepackage{adjustbox}
\usepackage{dsfont}
\RequirePackage{amsthm,amsmath,amsfonts,amssymb}
\RequirePackage[authoryear]{natbib}
\RequirePackage[colorlinks,citecolor=blue,urlcolor=blue]{hyperref}
\RequirePackage{graphicx}
\RequirePackage{multirow}
\RequirePackage{xcolor}
\RequirePackage{float}

\providecommand{\keywords}[1]
{
  \small	
  \textbf{\textit{Keywords---}} #1
}

\title{Targeted empirical Bayes for more supervised joint factor analysis}
\author{Glenn Palmer$^1$ and David B. Dunson$^{1,2}$}
\date{%
    $^1$Department of Statistical Science, Duke University\\%
    $^2$Department of Mathematics, Duke University\\[2ex]%
}

\begin{document}

\maketitle

\begin{abstract}
Joint Bayesian factor models are popular for characterizing relationships between multivariate correlated predictors and a response variable. Standard models assume that all variables, including both the predictors and the response, are conditionally independent given latent factors. In marginalizing out these factors, one obtains a low rank plus diagonal factorization for the joint covariance, which implies a linear regression for the response given the predictors. Although there are many desirable properties of such models, these methods can struggle to identify the signal when the response is not dependent on the dominant principal components in the predictors. To address this problem, we propose estimating the residual variance in the response model with an empirical Bayes procedure that targets predictive performance of the response given the predictors. We illustrate that this can lead to substantial improvements in simulation performance. We are particularly motivated by studies assessing the health effects of environmental exposures and provide an illustrative application to NHANES data. 
\end{abstract}

\keywords{Collinearity; Factor analysis; targeted empirical Bayes; High-dimensional predictors; Joint model; Mixture exposures; Principal components regression}

\section{Introduction}

Modern scientific data sets often involve both high-dimensional and highly correlated observed traits, making a simple linear regression analysis, for example via ordinary least squares, infeasible. To address this challenge, two broad categories of more sophisticated linear models have emerged over the past few decades. The first category assumes that the regression coefficients are sparse, which means some fraction of them are zero. Popular sparsity-inducing methods include the Lasso \citep{tibshirani1996regression} and the elastic net \citep{zou2005regularization}, among many others. Alternatively, to avoid making the sparsity assumption in situations where it may be inappropriate, one can instead assume that the $p$ predictors share some sort of lower-dimensional structure that can be exploited. Classical approaches in this category include principal component regression \citep{massy1965principal} and joint latent factor models \citep{west2003bayesian}, which are the focus of this work.

Joint latent factor models work by assuming that predictors $x_i \in \mathbb{R}^p$ and response $y_i \in \mathbb{R}$ can be expressed as functions of latent variables $\eta_i \in \mathbb{R}^k$ plus noise, where $k < p+1$. When the functions are linear and $\eta_i$ and the noise are given Gaussian distributions, the joint distribution of $(x_i^T, y_i)^T$ can be used to compute the linear regression coefficients for $y_i | x_i$. A key challenge is choosing $k$ to balance parsimony and flexibility. Traditionally, evaluation metrics such as AIC or BIC are used; however, in recent years, overfitted factor models have been developed that use shrinkage priors to remove unnecessary factors
\citep{bhattacharya2011sparse, legramanti2020bayesian, fruhwirth2023generalized}. Despite the convenience and inferential benefits of these approaches, a downside in the regression setting is that the likelihood used to choose $k$ is the $(p+1)$-dimensional joint likelihood of $(x_i^T,y_i)^T$. If $p$ is large, predictors $x_i$ can dominate this likelihood, leading to an underestimate of the importance of latent factors that are crucial for the conditional distribution $y_i | x_i$, but that explain a relatively small fraction of the overall variation in $(x_i^T,y_i)^T$ \citep{hahn2013partial}.

In this work, we propose a Targeted Empirical Bayes Factor Regression (TEB-FAR) approach to solve this problem. TEB-FAR chooses the residual variance in the response component to minimize the error in predicting $y_i$ from $x_i$. We use the terminology {\em targeted} to emphasize that we are not estimating the hyperparameter to maximize the joint likelihood, following common practice in the empirical Bayes literature, but instead focus on the conditional distribution of $y_i|x_i$. In the remainder of this section, we define joint factor models and summarize some related literature on dimensionality reduction in the supervised regression setting. In Section \ref{our_approach}, we motivate and describe our TEB-FAR approach. In Section \ref{results}, we present results that compare our approach to competitors in both simulations and applications, with particular emphasis on studying the relationship between chemical exposures and health outcomes. Finally, in Section \ref{discussion}, we conclude with a discussion of our findings and some directions for future work.

\subsection{Latent factor models}\label{factor_regression}

Gaussian joint factor models induce a regression of $y_i \in \mathbb{R}$ on $x_i \in \mathbb{R}^p$ by letting
\begin{equation}\label{factor_model}
    (x_i^T,y_i)^T \sim N_p(\Lambda \eta_i, \Sigma), \,\,\,\,\,\, \eta_i \sim N_k(0, I) \,\,\,\,\,\, \text{ for } i=1,\ldots,n,
\end{equation}
where the data are centered prior to analysis, $\Lambda \in \mathbb{R}^{(p+1) \times k}$ is the \textit{factor loadings matrix}, $\eta_1,\ldots,\eta_n \in \mathbb{R}^k$ are \textit{latent factors}, $\Sigma = \text{diag}(\sigma_1^2, \ldots, \sigma_p^2, \sigma_y^2)$ is a diagonal matrix of \textit{idiosyncratic variances}, and $k < p+1$. We induce the distribution
$(x_1^T,y_1)^T,\ldots,(x_n^T,y_n)^T \overset{i.i.d.}{\sim} N_p(0, \Lambda \Lambda^T + \Sigma)$,
which can be used to compute the conditional distribution
$y_i | x_i \sim N(\beta^T x_i, \sigma^2),$
where $\beta^T \in \mathbb{R}^p$ and $\sigma^2$ are functions of the covariance matrix $\Lambda \Lambda^T + \Sigma$. Thus, we induce a form of regularized linear regression, in which a smaller $k$ results in greater regularization. For an introduction to factor analysis, see \cite{bartholomew2011latent}.

Joint factor models use the marginal covariance in $x$ to inform the estimation of latent factors, which potentially provides advantages over methods that focus only on the conditional of $y|x$. However, this also leads to some disadvantages: (1) methods for inferring $k$ tend to find the dominant factors underlying the covariance in $x$ and may miss more subtle factors highly related to $y$, and (2) the inferred factors depend substantially more on $x$ than on $y$. Both of these issues can degrade predictive performance and inferences in the induced $y|x$ model, problems that we propose to address through a simple modification.

\subsection{Related literature}

Our problem of supervised factor analysis is related to principal component regression. The classical approach, illustrated by \cite{jeffers1967two}, for example, is to perform a principal component analysis on $X$, and then to keep the top $k$ components for use in regression based on their associated eigenvalues. Although this approach is still common in practice, a number of authors have criticized it. \cite{jolliffe1982note} presents several real-world examples in which later components that would be discarded are important for explaining an outcome. \cite{cox1968notes} notes that ``there is no logical reason why the dependent variable should not be closely related to the least important principal component.'' \cite{hadi1998some} illustrate this phenomenon with a data set where the first $p-1$ components explain $99.96\%$ of the variation in $X$, but the final component is the only one related to the outcome. To address these issues, several authors have suggested alternative methods of variable selection among the principal components, regardless of the amount of variation they explain in $X$ \citep{cox1968notes, jolliffe1982note, sutter1992principal, pires2008selection}. As an alternative, \cite{bair2006prediction} developed a supervised PCA method that performs variable selection \textit{before} computing the principal components. They compute univariate regression coefficients for the outcome on each predictor separately and then perform PCA on the matrix of only those predictors whose coefficients are above a threshold. Related supervised approaches have been developed for the probabilistic \citep{yu2006supervised}, sparse \citep{sharifzadeh2017sparse}, and functional \citep{nie2018supervised} PCA settings.

In the factor analysis literature, several methods have been proposed to choose the number of factors $k$. In the Bayesian paradigm, these include treating $k$ as a model parameter and sampling it using reversible jump MCMC \citep{green1995reversible, lopes2004bayesian}, or alternately taking an over-fitted factor model approach relying on priors that shrink unnecessary columns of $\Lambda$ to zero \citep{bhattacharya2011sparse, legramanti2020bayesian, fruhwirth2023generalized}. However, since these methods select $k$ and estimate all model parameters using the joint likelihood of outcomes $y_1,\ldots,y_n \in \mathbb{R}$ and predictors $x_1,\ldots,x_n \in \mathbb{R}^p$, predictors have been found to dominate factor selection, particularly when $p$ is large \citep{hahn2013partial}. To address this, \cite{hahn2013partial} propose a more general regression factor model that relaxes the conditional independence assumption of $y_i\perp\kern-5pt\perp x_i \mid \eta_i$, instead allowing additional dependence directly between $y_i$ and $x_i$. Similarly, \cite{fan2024latent} suggest an approach that models $y_i$ as a function of both latent factors $\eta_i$ and predictor idiosyncratic errors, $\varepsilon_{ix}$. In both of these approaches, the authors assume that the latent factor model is insufficient and thus add additional dependence between $y_i$ and $x_i$. In contrast, we seek to maintain the simple factor model setup, but to estimate better factors and loadings for modeling $y_i | x_i$.

\section{Our approach}\label{our_approach}

\subsection{Motivation}\label{motivation}

To motivate our approach, we consider an extension of the motivating example from \cite{hahn2013partial}. Suppose data $(x_i^T,y_i)^T$ are generated from the latent factor model in \eqref{factor_model} with
$$\Lambda^T = \begin{bmatrix}
    0 & -4 & 0 & -8 & -4 & -6 & 1 & -1 & 4 & 0 \\
    1 & 0 & 0 & -1 & 0 & 1 & 0 & 1 & 0 & 1
\end{bmatrix},
\,\,\,\,\,\,\,\, \Sigma=\text{diag}(0.2,\ldots,0.2).$$
If we assume the data come from a mean-zero Gaussian factor model, but with $k$ underestimated to be $1$ instead of $2$, the closest $1$-factor model to the true model in terms of KL divergence is defined by
$$\Tilde{\Lambda}^T = \begin{bmatrix}
    0.0004 & -4.0016 & -0.0000 & -7.9810 & -4.0016 & -5.9853 & 1.0002 & -0.9973 & 4.0016 & 0.0004
\end{bmatrix},$$ 
$$\Tilde{\Sigma}=\text{diag}(1.2000, 0.1872, 0.2000, 1.5032, 0.1872, 1.3763, 0.1996, 1.2054, 0.1872, 1.2000).$$
Observe that the single column of $\Tilde{\Lambda}$ is almost identical to the first column of $\Lambda$ from the full model. Although this $1$ factor approximation explains most of the variation in the joint distribution, it is totally useless if we wanted to predict $y_i$ (the final index) using $x_i$ (the first nine indices); it effectively models $y_i$ as mean zero noise independent of the other variables \citep{hahn2013partial}.

\begin{figure}[h]
\centering
\includegraphics[width=0.75\textwidth]{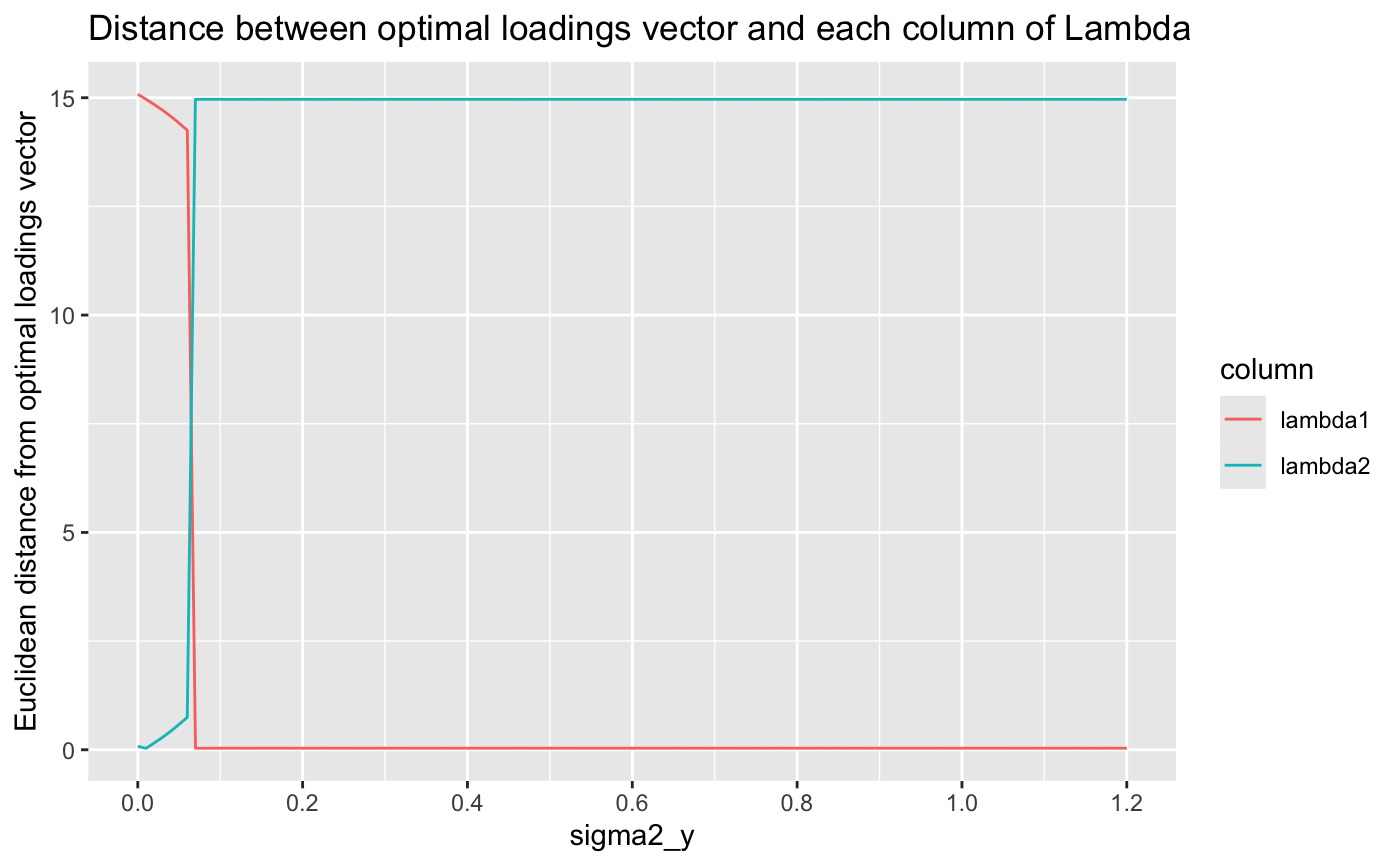}
\caption{\small{Distance between the KL-optimal loadings vector in the 1-factor model and each column of the true loadings matrix $\Lambda$ as a function of fixed $\sigma^2_y$.}}
\label{fig_phase_transition}
\end{figure}

We would like some way to encode in our model a strengthened preference for learning $y_i|x_i$, so that (1) it avoids making such a fatal underestimation of $k$, and (2) for a given $k$, it estimates better factors for predicting $y_i$. To do so, we observe that in the true model above, the idiosyncratic variance term $\sigma^2_{y}$ is only $0.2$, while in the $1$ factor model, it is estimated as $1.2$, the marginal variance of $y_i$. This observation suggests a question: Can forcing $\sigma^2_y$ to a smaller value change the emphasis the model places on the two true factors? To answer this question, we examine how the optimal $1$-factor approximation to the true model changes when the optimization is performed conditional on a fixed value of $\sigma^2_y$. The squared Euclidean distances from the KL-optimal single factor loadings vector to each of $\lambda_1$ and $\lambda_2$ (the first and second columns of $\Lambda$) are shown in Figure \ref{fig_phase_transition} along a grid of $\sigma^2_y$ values between $0$ and $1.2$. 
Observe that at approximately $\sigma^2_y = 0.06$, the optimal loadings vector undergoes a sort of phase transition from being essentially equal to $\lambda_1$ to essentially equal to $\lambda_2$. Thus, by shrinking $\sigma^2_y$ closer and closer to $0$, we can in some sense ``reorder'' the factors in terms of the priority they are given by the joint likelihood. Moreover, even at higher values of $\sigma^2_y$ than are needed to cause the phase transition, the second factor is becoming more and more important to the joint likelihood as $\sigma^2_y$ becomes smaller, which should push models to choose $k$ to be high enough so that it is included.

\subsection{Our approach}

Given the motivation above, we define our TEB-FAR model as the factor model of \eqref{factor_model}, but with an important alteration: Instead of inferring the joint posterior of $\sigma_y^2$ and the other model parameters using a fully Bayesian approach, we propose an empirical Bayesian approach to estimate $\hat{\sigma}_y^2$ as the value optimizing predictive performance of $y$ from $x$. In particular, for standardized data such that $y_i$ has marginal empirical mean $0$ and variance $1$, $\hat{\sigma}_y^2$ is selected by cross-validation from values on a grid from $0$ to $1$. Then, fixing $\hat{\sigma}_y^2$, we use the increasing shrinkage prior of \cite{bhattacharya2011sparse} to choose the number of factors, $k$. Specifically, this setup chooses a large upper bound $\Tilde{k}$ for $k$, and then models elements of $\Lambda$ with the priors
$$\lambda_{jl} \sim N(0, \xi_{jl}^{-1} \tau_{l}^{-1}), \,\,\,\,\,\,\,\,\,\,\,\, j=1,\ldots,p; \,\,\, l=1,\ldots,\Tilde{k}.$$
In the above, the $\xi_{jl}$s are assigned independent gamma prior distributions, while the $\tau_l$s are modeled as a product of gamma random variables, specified so that $\tau_l^{-1}$ decreases for increasing $l$ at a rate learned from the data. Thus, for high values of $l$, the entries of $\Lambda$ are effectively shrunk to zero. The full model setup with all prior distributions is specified in the Supplementary Information.

\subsection{Mathematical intuition}

Building on the motivation from Section \ref{motivation}, here we examine the induced regression coefficients in a one-factor setup. Let $\gamma \in \mathbb{R}$ be the factor loading for $y_i$, so that
\begin{align*}
    y_i &= \gamma \eta_i + \varepsilon_{yi}, \,\,\,\,\,\,\,\,\,\, \varepsilon_{yi} \sim N(0, \sigma_y^2),\\
    x_i &= \lambda \eta_i + \varepsilon_{xi}, \,\,\,\,\,\,\,\,\,\, \varepsilon_{xi} \sim N(0, \Sigma),\\
    \eta_i &\sim N(0,1), \,\,\,\,\,\,\,\,\,\,\,\,\,\, \Sigma = \text{diag}(\sigma_1^2,\ldots,\sigma_p^2).
\end{align*}
Then the induced linear regression of $y_i$ on $x_i$ can be written as $y_i | x_i \sim N(\beta^T x_i, \sigma^2)$, where
$$\beta = \frac{\gamma}{1 + \sum_{j=1}^p \frac{\lambda_j^2}{\sigma_j^2}} \left(\frac{\lambda_1}{\sigma_1^2},\ldots,\frac{\lambda_p}{\sigma_p^2} \right)^T\quad \mbox{and}\quad \sigma^2 = \sigma_y^2 + \frac{\gamma^2}{1 + \sum_{j=1}^p \frac{\lambda_j^2}{\sigma_j^2}}.$$
Observe in the variance formula above that if we force $\sigma_y^2$ to be arbitrarily small, the model will eventually be forced to compensate by increasing $\gamma^2$ away from zero. But then, in doing so, the magnitude of the regression coefficients $\beta$ will also be inflated, increasing the penalty for estimating a loadings vector $\lambda$ that is suboptimal for predicting $y_i | x_i$. Thus, by forcing $\sigma_y^2$ to be small, we can force the 1-factor model to find signal for $y_i |x_i$ when it is present, even at a cost to the likelihood for $x_i$.

\begin{figure}[H]
\centering
\includegraphics[width=0.75\textwidth]{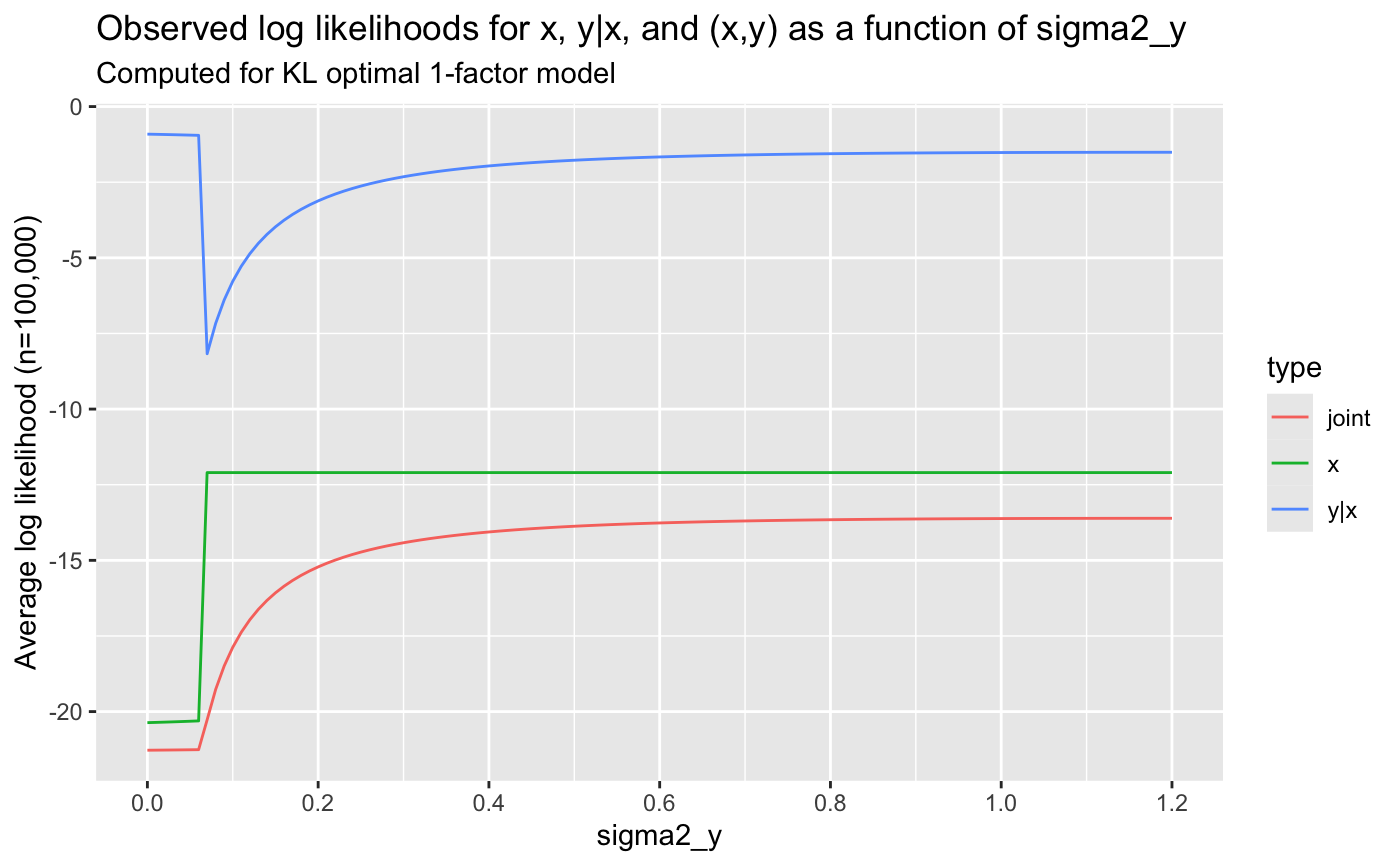}
\caption{\small{Log likelihoods for the KL-optimal 1-factor models conditional on fixed values of $\sigma_y^2$, computed for 100,000 observations generated from the true 2-factor model described in Section \ref{motivation}. By definition, the joint log likelihood for $(x,y)$ is equal to the sum of the log likelihoods for $x$ and $y|x$.}}
\label{fig_phase_transition_likelihoods}
\end{figure}

To illustrate this, Figure \ref{fig_phase_transition_likelihoods} shows the 1-factor log likelihoods for 100,000 observations generated from the true 2-factor model from Section \ref{motivation}, computed for the KL-optimal 1-factor models on a grid of $\sigma_y^2$ values. Observe that as $\sigma_y^2$ is decreased toward $0.06$, the log likelihood for $y_i | x_i$ suffers dramatically. Eventually, the improvement to this conditional likelihood from switching to the second true factor exceeds the cost to the joint likelihood of $x_i$, and the phase transition occurs. The joint model for $(x_i^T,y_i)^T$ is worse than the naive model that learns the first factor, but if our goal is to predict $y_i$ using $x_i$, we have improved our model dramatically.

Although the formulas here are for the 1-factor case, we postulate that this intuition generalizes for $k > 1$, and we support this hypothesis with empirical results in Section \ref{results}.


\section{Simulation and data analysis results}\label{results}

\subsection{NHANES phthalate and BMI data}

\begin{figure}[h]
\centering
\includegraphics[width=\textwidth]{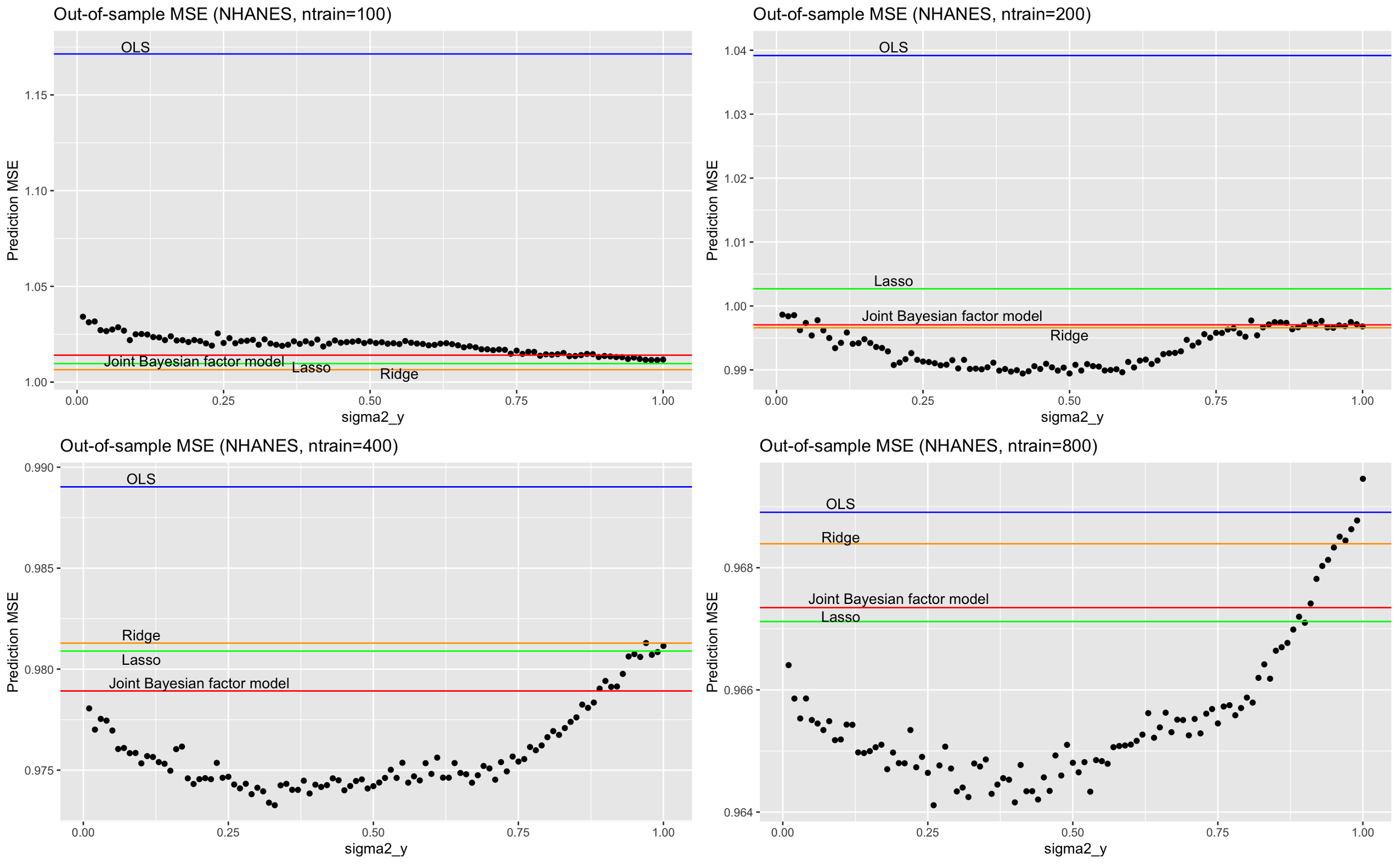}
\caption{\small{Average out-of-sample prediction MSE for the NHANES data set across $50$ train/test splits as a function of fixed $\sigma^2_y$ for four different training set sizes. Lasso and ridge MSE values are for the optimal choice of their tuning parameters over a grid.}}
\label{fig_NHANES_4panel}
\end{figure}

To evaluate our empirical Bayes approach for estimating $\sigma_y^2$, we applied it to predict patient BMI using measured exposure levels to $p=19$ phthalates, using data from the 2017-2018 NHANES public repository, available at https://wwwn.cdc.gov/nchs/nhanes/. For simplicity, we considered a complete case analysis of only adults over the age of 18, resulting in a sample of $n=1724$ participants. Given these data, we randomly generated $50$ partitions of the sample into training and test sets, for each of a range of training set sizes. After standardizing the data, we fit our TEB-FAR model to each training set on a grid of $100$ $\sigma_y^2$ values from $0.01$ to $1$, and used the fitted models to predict the BMI for the test set. For comparison, we also fit a joint Bayesian factor model using a multiplicative gamma process prior for the loadings matrix as implemented in the ``infinitefactor'' R package \citep{poworoznek2020package}, lasso and ridge regression as implemented in the ``glmnet'' R package \citep{friedman2021package}, each for a grid of tuning parameter values and taking the best one, and OLS linear regression. The average prediction MSE for each model is shown in Figure \ref{fig_NHANES_4panel} for training set sizes of $100$, $200$, $400$, and $800$. Additional results for $ntrain=50$ and $1600$, as well as code to reproduce these results, are contained in the supplement. The results for $ntrain=50$ and $1600$ are qualitatively similar to those for $ntrain=100$ and $800$, respectively.

In Figure \ref{fig_NHANES_4panel}, observe that for $ntrain=100$, all methods have MSE greater than $1$, indicating that they would all be outperformed by simply predicting the mean. However, for $ntrain=200$, $400$, and $800$, TEB-FAR outperforms all competitors for a substantial range of $\sigma_y^2$ values. Interestingly, the average posterior mean $\sigma_y^2$ value estimated by the joint Bayesian factor model across the different sizes of training sets is $0.90$, $0.90$, $0.86$, and $0.86$, suggesting that even as more training data are included, the inferred values of $\sigma_y^2$ remain very similar, despite the higher predictive accuracy that is possible if $\sigma_y^2$ is reduced to $\sim0.5$.

\subsubsection{Estimated factors by TEB-FAR vs. joint Bayesian factor model}

\begin{figure}[H]
\centering
\includegraphics[width=\textwidth]{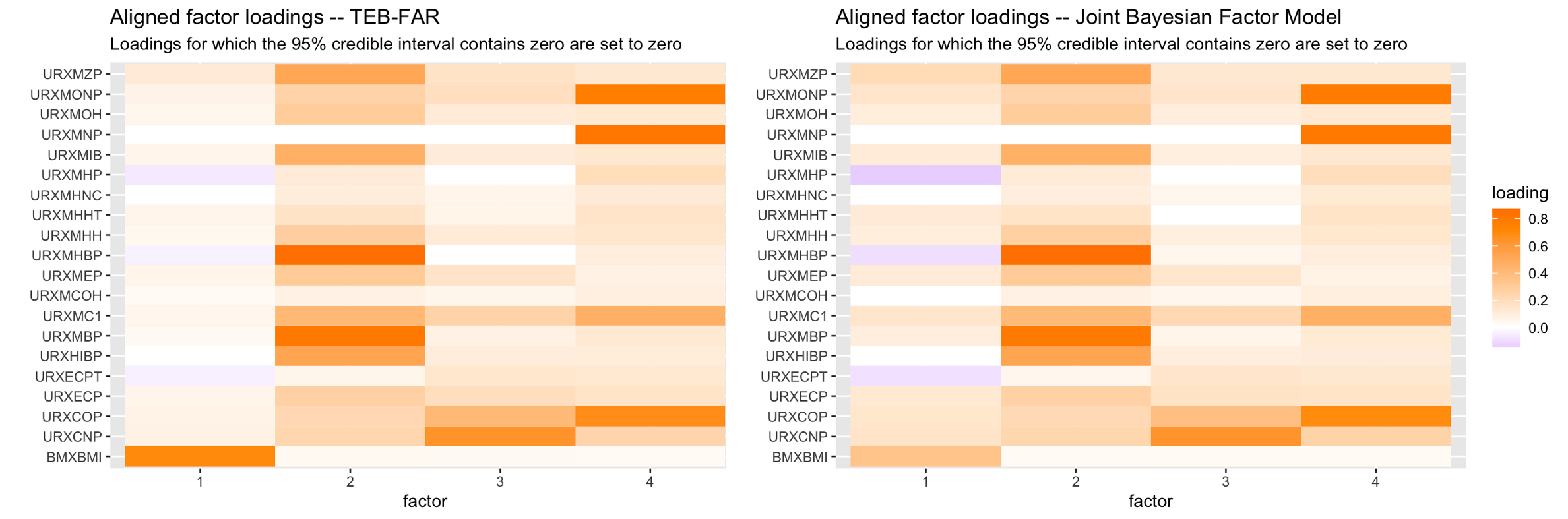}
\caption{\small{Aligned posterior mean columns of $\Lambda$ with the four highest outcome loading magnitudes estimated by TEB-FAR with $\sigma^2_y = 0.5$ (left) and a joint Bayesian factor model with multiplicative gamma process prior on $\Lambda$ (right). In both panels, these columns include all aligned columns $j$ for which $|\lambda_{y,j}| > 0.01$. The loadings for the outcome $y_i$ (BMXBMI) are in the bottom row. Entries for which the posterior $95\%$ credible interval includes zero are set to zero. }}
\label{fig_aligned_factors_NHANES}
\end{figure}

To understand the shifts in parameter estimates driving our approach's gains in predictive performance, we fit our TEB-FAR model with $\sigma_y^2 = 0.5$, which is within the range that outperforms competitors for $ntrain \geq 200$, as well as a joint Bayesian factor model (JBFM) with a multiplicative gamma process prior to the loadings matrix to the entire NHANES data set ($n=1724$). Since factors and loadings are not identifiable, we then applied the approach of \cite{poworoznek2021efficiently} to align the columns of $\Lambda$ in each model. In doing so, Figure \ref{fig_aligned_factors_NHANES} shows the posterior mean of the aligned columns for each model that include a loading of at least $0.01$ for $y_i$ (BMI). Observe that the second, third, and fourth columns in each panel are almost identical between the models. However, the first column, which has by far the largest loading for $y_i$, differs substantially. In particular, the posterior mean loading for $y_i$ roughly doubles from $0.35$ in the JBFM to $0.70$ in TEB-FAR, corresponding to a shift from explaining $\sim 12\%$ of the standardized outcome variance to $\sim 49\%$; this corresponds closely to the decrease of $\sigma_y^2$ from the posterior mean of the JBFM of $0.88$ to $0.5$ in TEB-FAR. The remaining entries in this column are roughly halved, but with some meaningful discrepancies. For example, the loading for MBP decreased from $0.096$ to $0.035$, and MC1 decreased from $0.152$ to $0.054$, while the loading for MCOH went from having a $95\%$ credible interval that included zero to one that is strictly positive in TEB-FAR, with a posterior mean of $0.032$. The exact values of the posterior means for all the entries in Figure \ref{fig_aligned_factors_NHANES} can be found in the supplement, along with the induced covariances between BMI and each phthalate. Interestingly, the induced covariance matrix for TEB-FAR is much closer to the Pearson sample covariance matrix than the JBFM covariance, with the sum of squared differences in the entries dropping more than an order of magnitude from $5.6 \times 10^{-2}$ between the JBFM covariance and the sample covariance to $4.8 \times 10^{-3}$ between TEB-FAR and the sample covariance. This suggests that one interpretation of the differences between the models is simply a decrease in regularization when estimating the joint covariance.

\begin{figure}[H]
\centering
\includegraphics[width=\textwidth]{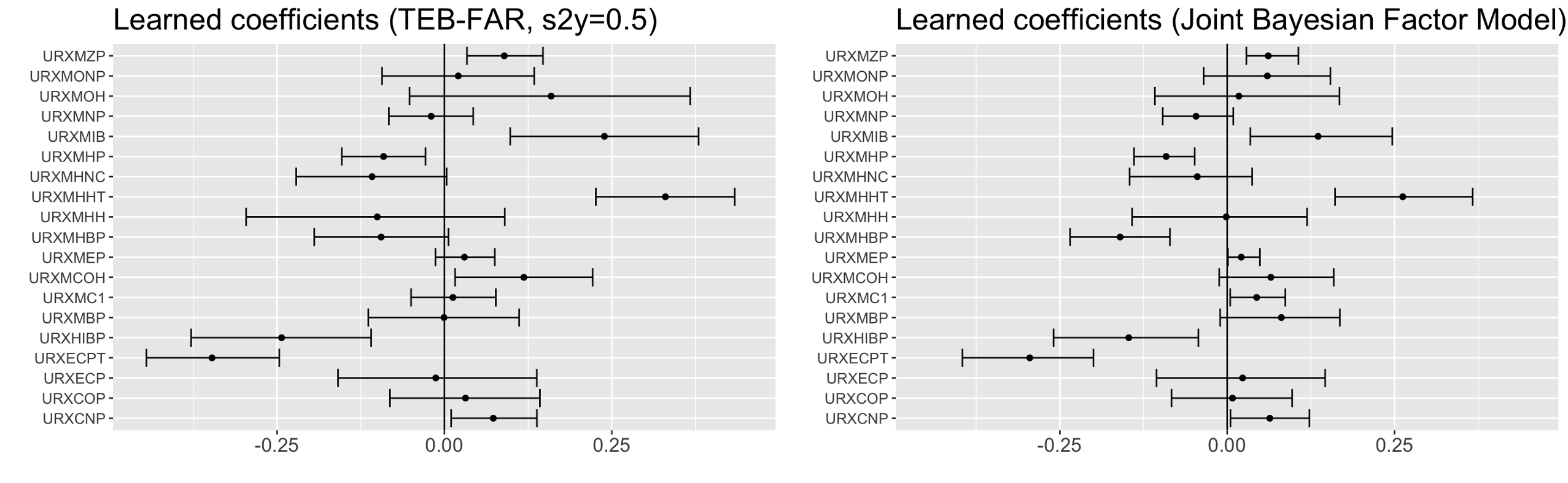}
\caption{\small{Posterior means and $95\%$ credible intervals for induced linear regression coefficients for BMI regressed on 19 phthalates using TEB-FAR with $\sigma^2_y = 0.5$ (left) and a joint Bayesian factor model with multiplicative gamma process prior on $\Lambda$ (right).}}
\label{fig_coefficients}
\end{figure}

Finally, in Figure \ref{fig_coefficients} we examine how these shifts in inferred model parameters affect the induced linear regression coefficients for $y_i$ on $x_i$. Qualitatively, the intervals are wider on average for TEB-FAR compared to the JBFM, and the magnitudes of the largest coefficients are greater. However, the shifts vary among the chemicals; estimated coefficients for MBP and MC1, which had the larger-than-average drops in loadings discussed above, both drop close to zero, with the $95\%$ credible interval for MC1 including zero, while it was strictly positive for the JBFM. On the other hand, the interval for MCOH is strictly positive for TEB-FAR but included zero for the JBFM, again mirroring the shift in factor loadings. Clearly, TEB-FAR produces a meaningfully different set of inferential conclusions relative to the joint Bayesian factor model. Given the corresponding improved predictive performance, it is reasonable to take these differences seriously.

\subsection{Stability of estimating $\sigma_y^2$}

\subsubsection{NHANES}

To verify that we can consistently estimate an appropriate $\sigma_y^2$ value for TEB-FAR, we again fit our model to the NHANES data, but instead of simply fitting all models on a grid of $\sigma_y^2$ values, we used 10-fold cross validation to estimate the emprical Bayes $\sigma_y^2$ value within each training set. We then used that value to train the model used to predict the values of the test set. We again considered the same competitors; for lasso and ridge, we selected $\lambda$ values using the built-in cross-validation function in the glmnet package \citep{friedman2021package}. The MSE values of the test set for these predictions are shown in Table 1, averaged over $50$ random train/test splits in each row. Similarly to Figure \ref{fig_NHANES_4panel}, the joint Bayesian factor model outperforms TEB-FAR for $ntrain \leq 100$, and TEB-FAR outperforms all competitors for $ntrain \geq 400$. For $ntrain=200$, TEB-FAR and JBFM are quite close in performance, with JBFM having slightly lower MSE, but TEB-FAR having lower MSE in $28/50$ splits. Interestingly, the performance of lasso and ridge appears to suffer substantially now that their tuning parameter is selected within each training set. For $ntrain=400$, in particular, the MSE gap between these models and the factor models is much larger than in Figure \ref{fig_NHANES_4panel}, and they actually perform slightly worse than OLS regression. This suggests that lasso and ridge may suffer from more instability in tuning parameter selection than TEB-FAR does in its empirical Bayes estimation of $\sigma_y^2$. An explanation for this is that TEB-FAR is estimating $\sigma_y^2$ from $[0,1]$ for standardized data, while such a restriction does not make sense for the parameters in lasso and ridge.

\begin{table}[h]
\centering
\caption{\small{Test-set prediction MSE for TEB-FAR and four competitors for NHANES data, averaged over 50 random train/test splits for each training set size. For TEB-FAR, the optimal $\sigma_y^2$ value was chosen using 10-fold cross-validation within each training set. Bold values indicate the smallest MSE in each row.}}
\begin{tabular}{llllll}
\hline
            & TEB-FAR & Joint Bayesian Factor Model & Lasso & Ridge & OLS   \\ \hline
ntrain=50   & 1.044    & \textbf{1.029}                & 1.035 & 1.035 & 1.589 \\
ntrain=100  & 1.022    & \textbf{1.016}                & 1.025 & 1.019 & 1.173 \\
ntrain=200  & 0.993    & \textbf{0.992}                & 1.006 & 1.003 & 1.034 \\
ntrain=400  & \textbf{0.973}    & 0.976                & 0.986 & 0.987 & 0.985 \\
ntrain=800  & \textbf{0.955}    & 0.956                & 0.958 & 0.960 & 0.959 \\
ntrain=1600 & \textbf{0.931}    & 0.933                & 0.933 & 0.933 & 0.932 \\ \hline
\end{tabular}
\end{table}

\subsubsection{Simulations}

Finally, we evaluated our approach for three simulation scenarios across a range of training set sizes. In Scenario 1, we generate data from a factor model with $p=20$ and $k=10$. To generate the loadings for $x_i$, we randomly selected $10$ entries in each column to be nonzero, and then generated those entries as i.i.d. $\text{Exponential}(1)$. We then scaled the columns so that they had Euclidean norms $1, 0.9, \ldots, 0.1$. Then, for all replications, we set the loadings for $y_i$ to $(0,\ldots,0,1)$, so that the outcome only loads on the factor that explains the least amount of variance in $x_i$. We let all indiosyncratic variances be $0.2$, rescaled the resulting covariance matrix so that all marginal variances were $1$, and then generated $(x_i^T, y_i)^T \in \mathbb{R}^{21}$ as multivariate Gaussian. In Scenario 2, we generated data the same way as in Scenario 1, except that we generated the loadings for $y_i$ as part of the method we described for generating the $x_i$ loadings, so that the outcome does not only load on the final factor. In general, this should lead to higher signal-to-noise ratios. In Scenario 3, we generate $x_i$ from a factor model with $p=20$, $k=8$, all entries of $\Lambda$ generated as i.i.d. $\text{Exponential}(1)$, and $\Sigma = \text{diag}(0.01,\ldots,0.01)$. To generate $y_i$, we randomly chose $12$ nonzero indices from $x_i$ and generated linear regression coefficients from $N(0,1)$, so that the true linear regression equation has sparse coefficients. We then scaled $\text{Cov}(x_i)$ and the regression coefficients so that all variables have marginal variance $1$ and the regression variance is $0.9$. The outcome was generated as normally distributed with mean $\beta^T x_i$ and variance $0.9$, so the true regression $R^2$ is $0.1$ for all replications. For all scenarios and sizes of the training set, we repeated the data generation process for $50$ random initializations, and in each replication we computed the MSE of the test set for the predictions on a test set of $ntest=\text{100,000}$.

\begin{table}[h]
\centering
\caption{\small{Test-set prediction MSE for TEB-FAR and four competitors for simulated data sets, averaged over 50 random seeds. Optimal $\sigma_y^2$ values were estimated for each seed using 10-fold cross-validation within the training set. Bold values indicate the smallest MSE in each row.}}
\resizebox{\textwidth}{!}{%
\begin{tabular}{llllll}
\hline
            & TEB-FAR & Joint Bayesian Factor Model & Lasso & Ridge & OLS   \\ \hline
Simulation Scenario 1 (ntrain=200)          & 1.010         & 1.007                     & 1.008      & \textbf{1.006}      & 1.076    \\
Simulation Scenario 1 (ntrain=500)          & 0.998         & 0.997                     & 0.996      & \textbf{0.995}      & 1.015    \\
Simulation Scenario 1 (ntrain=1000)          & 0.987         & 0.990                     & \textbf{0.987}      & 0.988      & 0.993    \\
Simulation Scenario 1 (ntrain=1500)          & \textbf{0.983}         & 0.985                     & 0.984      & 0.984      & 0.986    \\
Simulation Scenario 1 (ntrain=2000)          & \textbf{0.981}         & 0.982                     & 0.982      & 0.982      & 0.982    \\
\hline
Simulation Scenario 2 (ntrain=200)          & 0.832         & \textbf{0.832}                     & 0.846      & 0.846      & 0.880    \\
Simulation Scenario 2 (ntrain=500)           & 0.813         & \textbf{0.813}                     & 0.821      & 0.822      & 0.830    \\
Simulation Scenario 2 (ntrain=1000)          & 0.805         & \textbf{0.804}                     & 0.809      & 0.810      & 0.812    \\
Simulation Scenario 2 (ntrain=1500)          & 0.802         & \textbf{0.801}                     & 0.805      & 0.805      & 0.805    \\
Simulation Scenario 2 (ntrain=2000)          & 0.800         & \textbf{0.800}                     & 0.803      & 0.803      & 0.803    \\
\hline
Simulation Scenario 3 (ntrain=200)          & 0.928         & \textbf{0.927}     & 0.939      & 0.930      & 0.992   \\
Simulation Scenario 3 (ntrain=500)          &  0.916        & \textbf{0.915}                     & 0.918      & 0.916      & 0.939    \\
Simulation Scenario 3 (ntrain=1000)          & 0.908         & \textbf{0.908}                     & 0.909      & 0.909      & 0.918    \\
Simulation Scenario 3 (ntrain=1500)          & 0.906         & 0.906                     & 0.907      & \textbf{0.906}      & 0.913    \\
Simulation Scenario 3 (ntrain=2000)          & 0.905         & 0.905                     & 0.906      & \textbf{0.905}      & 0.910    \\
\hline
\end{tabular}}
\label{table2}
\end{table}

The results of these simulations are shown in Table \ref{table2}. For Scenario 1, observe that for small training sets, TEB-FAR is outperformed by the joint Bayesian factor model, as well as lasso and ridge. However, for $ntrain \geq 1500$, our empirical Bayes approach for estimating $\sigma_y^2$ allows the model to find additional signal missed by the other approaches, leading to the best predictive MSE of all models considered. This transition to superior performance above a threshold training set size mirrors the results we observed for the NHANES data. In Scenarios 2 and 3, TEB-FAR performs similarly to the joint Bayesian factor model across the board. This is perhaps unsurprising for Scenario 2, and suggests that when the true data generation mechanism is a Gaussian factor model with high signal to noise ratio, there may not be much to gain from our procedure for estimating $\sigma_y^2$, and using it may actually hurt performance slightly. Similarly, when the outcome regression model is sparse and does not arise from a factor model, as is the case in Scenario 3, TEB-FAR may again not allow one to find additional signal. 

\section{Discussion}\label{discussion}

In this work, we proposed a simple modification to Bayesian joint Gaussian factor models to make them ``more supervised'' to address the common problem that inferred factors can essentially ignore the response, leading to poor predictive inferences.  Our
key idea was to infer the outcome residual variance, $\sigma_y^2$, by an empirical Bayes procedure targeting  predictive performance of $y$ given $x$ instead of using a fully Bayes procedure targetting the joint of $y$ and $x$ for all parameters.
We demonstrated with both NHANES chemical exposure data and simulations that our approach can improve predictive performance relative to competitors. We also showed that the shifts in the model corresponding to these predictive gains have meaningful implications for inference, which may be of interest to scientists.

There are several promising directions for future work. One avenue is to generalize this idea of targeted empirical Bayes to other settings beyond Gaussian joint factor models. There are likely many situations in which simply taking a fully Bayesian approach for all the parameters can have unappealing consequences when the analysis goals are also targeted; e.g. when certain components of the likelihood can be considered as a nuisance. A second interesting direction is to consider other approaches to make joint latent factor models more supervised. Although there is a substantial literature on algorithmic approaches to supervised PCA, much less consideration has been given to Bayesian approaches. An alternative to the targeted empirical Bayes approach proposed in this article is to take a Bayesian decision-theoretic approach, perhaps choosing $k$ to be intentionally too high, but then post-processing the resulting model to identify a subset of relevant factors based on a loss function. Decision-theoretic post-processing of posterior samples has been shown to be effective in the sparse regression setup \citep{hahn2015decoupling, li2025bayesian} and for factor analysis to summarize sparse factor loadings based on the joint distribution \citep{bolfarine2024decoupling}, and it may be a promising direction for more supervised Bayesian factor analysis as well.


\section{Acknowledgments}
This work was supported by the National Institutes of Health (NIH) grant R01ES035625 and by Merck \& Co., Inc., through its support for the Merck BARDS Academic Collaboration. The authors also acknowledge the Duke Compute Cluster for computational time.

\section{Code availability}
Code to reproduce all results can be found at https://github.com/glennpalmer/TEB-FAR.


\bibliographystyle{plainnat}
\bibliography{ref}

\begin{thebibliography}{28}
\providecommand{\natexlab}[1]{#1}
\providecommand{\url}[1]{\texttt{#1}}
\expandafter\ifx\csname urlstyle\endcsname\relax
  \providecommand{\doi}[1]{doi: #1}\else
  \providecommand{\doi}{doi: \begingroup \urlstyle{rm}\Url}\fi

\bibitem[Bair et~al.(2006)Bair, Hastie, Paul, and Tibshirani]{bair2006prediction}
Eric Bair, Trevor Hastie, Debashis Paul, and Robert Tibshirani.
\newblock Prediction by supervised principal components.
\newblock \emph{Journal of the American Statistical Association}, 101\penalty0 (473):\penalty0 119--137, 2006.

\bibitem[Bartholomew et~al.(2011)Bartholomew, Knott, and Moustaki]{bartholomew2011latent}
David~J Bartholomew, Martin Knott, and Irini Moustaki.
\newblock \emph{Latent Variable Models and Factor Analysis: A Unified Approach}.
\newblock John Wiley \& Sons, 2011.

\bibitem[Bhattacharya and Dunson(2011)]{bhattacharya2011sparse}
Anirban Bhattacharya and David~B Dunson.
\newblock Sparse {Bayesian} infinite factor models.
\newblock \emph{Biometrika}, 98\penalty0 (2):\penalty0 291--306, 2011.

\bibitem[Bolfarine et~al.(2024)Bolfarine, Carvalho, Lopes, and Murray]{bolfarine2024decoupling}
Henrique Bolfarine, Carlos~M Carvalho, Hedibert~F Lopes, and Jared~S Murray.
\newblock Decoupling shrinkage and selection in {Gaussian} linear factor analysis.
\newblock \emph{Bayesian Analysis}, 19\penalty0 (1):\penalty0 181--203, 2024.

\bibitem[Cox(1968)]{cox1968notes}
David~Roxbee Cox.
\newblock Notes on some aspects of regression analysis.
\newblock \emph{Journal of the Royal Statistical Society Series A: Statistics in Society}, 131\penalty0 (3):\penalty0 265--279, 1968.

\bibitem[Fan et~al.(2024)Fan, Lou, and Yu]{fan2024latent}
Jianqing Fan, Zhipeng Lou, and Mengxin Yu.
\newblock Are latent factor regression and sparse regression adequate?
\newblock \emph{Journal of the American Statistical Association}, 119\penalty0 (546):\penalty0 1076--1088, 2024.

\bibitem[Friedman et~al.(2021)Friedman, Hastie, Tibshirani, Narasimhan, Tay, Simon, and Qian]{friedman2021package}
Jerome Friedman, Trevor Hastie, Rob Tibshirani, Balasubramanian Narasimhan, Kenneth Tay, Noah Simon, and Junyang Qian.
\newblock Package ‘glmnet’.
\newblock \emph{CRAN R Repositary}, 595, 2021.

\bibitem[Fr{\"u}hwirth-Schnatter(2023)]{fruhwirth2023generalized}
Sylvia Fr{\"u}hwirth-Schnatter.
\newblock Generalized cumulative shrinkage process priors with applications to sparse {Bayesian} factor analysis.
\newblock \emph{Philosophical Transactions of the Royal Society A}, 381\penalty0 (2247):\penalty0 20220148, 2023.

\bibitem[Green(1995)]{green1995reversible}
Peter~J Green.
\newblock Reversible jump {Markov} chain {Monte} {Carlo} computation and {Bayesian} model determination.
\newblock \emph{Biometrika}, 82\penalty0 (4):\penalty0 711--732, 1995.

\bibitem[Hadi and Ling(1998)]{hadi1998some}
Ali~S Hadi and Robert~F Ling.
\newblock Some cautionary notes on the use of principal components regression.
\newblock \emph{The American Statistician}, 52\penalty0 (1):\penalty0 15--19, 1998.

\bibitem[Hahn and Carvalho(2015)]{hahn2015decoupling}
P~Richard Hahn and Carlos~M Carvalho.
\newblock Decoupling shrinkage and selection in {Bayesian} linear models: a posterior summary perspective.
\newblock \emph{Journal of the American Statistical Association}, 110\penalty0 (509):\penalty0 435--448, 2015.

\bibitem[Hahn et~al.(2013)Hahn, Carvalho, and Mukherjee]{hahn2013partial}
P~Richard Hahn, Carlos~M Carvalho, and Sayan Mukherjee.
\newblock Partial factor modeling: predictor-dependent shrinkage for linear regression.
\newblock \emph{Journal of the American Statistical Association}, 108\penalty0 (503):\penalty0 999--1008, 2013.

\bibitem[Jeffers(1967)]{jeffers1967two}
John~NR Jeffers.
\newblock Two case studies in the application of principal component analysis.
\newblock \emph{Journal of the Royal Statistical Society Series C: Applied Statistics}, 16\penalty0 (3):\penalty0 225--236, 1967.

\bibitem[Jolliffe(1982)]{jolliffe1982note}
Ian~T Jolliffe.
\newblock A note on the use of principal components in regression.
\newblock \emph{Journal of the Royal Statistical Society Series C: Applied Statistics}, 31\penalty0 (3):\penalty0 300--303, 1982.

\bibitem[Legramanti et~al.(2020)Legramanti, Durante, and Dunson]{legramanti2020bayesian}
Sirio Legramanti, Daniele Durante, and David~B Dunson.
\newblock Bayesian cumulative shrinkage for infinite factorizations.
\newblock \emph{Biometrika}, 107\penalty0 (3):\penalty0 745--752, 2020.

\bibitem[Li et~al.(2025)Li, Tokdar, and Xu]{li2025bayesian}
Aihua Li, Surya~T Tokdar, and Jason Xu.
\newblock A {Bayesian} decision-theoretic approach to sparse estimation.
\newblock \emph{arXiv preprint arXiv:2502.00126}, 2025.

\bibitem[Lopes and West(2004)]{lopes2004bayesian}
Hedibert~Freitas Lopes and Mike West.
\newblock Bayesian model assessment in factor analysis.
\newblock \emph{Statistica Sinica}, pages 41--67, 2004.

\bibitem[Massy(1965)]{massy1965principal}
William~F Massy.
\newblock Principal components regression in exploratory statistical research.
\newblock \emph{Journal of the American Statistical Association}, 60\penalty0 (309):\penalty0 234--256, 1965.

\bibitem[Nie et~al.(2018)Nie, Wang, Liu, and Cao]{nie2018supervised}
Yunlong Nie, Liangliang Wang, Baisen Liu, and Jiguo Cao.
\newblock Supervised functional principal component analysis.
\newblock \emph{Statistics and Computing}, 28:\penalty0 713--723, 2018.

\bibitem[Pires et~al.(2008)Pires, Martins, Sousa, Alvim-Ferraz, and Pereira]{pires2008selection}
Jos{\'e} Carlos~M Pires, Fernando~Gomes Martins, SIV Sousa, Maria~CM Alvim-Ferraz, and MC~Pereira.
\newblock Selection and validation of parameters in multiple linear and principal component regressions.
\newblock \emph{Environmental Modelling \& Software}, 23\penalty0 (1):\penalty0 50--55, 2008.

\bibitem[Poworoznek(2020)]{poworoznek2020package}
Evan Poworoznek.
\newblock infinitefactor: {Bayesian} infinite factor models.
\newblock \emph{CRAN Repository}, 2020.
\newblock URL \url{https://cran.r-project.org/web/packages/infinitefactor}.
\newblock R package version 1.0.

\bibitem[Poworoznek et~al.(2021)Poworoznek, Anceschi, Ferrari, and Dunson]{poworoznek2021efficiently}
Evan Poworoznek, Niccolo Anceschi, Federico Ferrari, and David Dunson.
\newblock Efficiently resolving rotational ambiguity in {Bayesian} matrix sampling with matching.
\newblock \emph{arXiv preprint arXiv:2107.13783}, 2021.

\bibitem[Sharifzadeh et~al.(2017)Sharifzadeh, Ghodsi, Clemmensen, and Ersb{\o}ll]{sharifzadeh2017sparse}
Sara Sharifzadeh, Ali Ghodsi, Line~H Clemmensen, and Bjarne~K Ersb{\o}ll.
\newblock Sparse supervised principal component analysis ({SSPCA}) for dimension reduction and variable selection.
\newblock \emph{Engineering Applications of Artificial Intelligence}, 65:\penalty0 168--177, 2017.

\bibitem[Sutter et~al.(1992)Sutter, Kalivas, and Lang]{sutter1992principal}
Jon~M Sutter, John~H Kalivas, and Patrick~M Lang.
\newblock Which principal components to utilize for principal component regression.
\newblock \emph{Journal of chemometrics}, 6\penalty0 (4):\penalty0 217--225, 1992.

\bibitem[Tibshirani(1996)]{tibshirani1996regression}
Robert Tibshirani.
\newblock Regression shrinkage and selection via the lasso.
\newblock \emph{Journal of the Royal Statistical Society Series B: Statistical Methodology}, 58\penalty0 (1):\penalty0 267--288, 1996.

\bibitem[West(2003)]{west2003bayesian}
Mike West.
\newblock Bayesian factor regression models in the “large p, small n” paradigm.
\newblock \emph{Bayesian Statistics}, 2003.

\bibitem[Yu et~al.(2006)Yu, Yu, Tresp, Kriegel, and Wu]{yu2006supervised}
Shipeng Yu, Kai Yu, Volker Tresp, Hans-Peter Kriegel, and Mingrui Wu.
\newblock Supervised probabilistic principal component analysis.
\newblock In \emph{Proceedings of the 12th ACM SIGKDD international conference on Knowledge discovery and data mining}, pages 464--473, 2006.

\bibitem[Zou and Hastie(2005)]{zou2005regularization}
Hui Zou and Trevor Hastie.
\newblock Regularization and variable selection via the elastic net.
\newblock \emph{Journal of the Royal Statistical Society Series B: Statistical Methodology}, 67\penalty0 (2):\penalty0 301--320, 2005.

\end{thebibliography}

\end{document}


\maketitle

\section{Full TEB-FAR model and prior distributions}

Taking $\Tilde{k}$ to be an upper bound on the number of factors and $\hat{\sigma}_y^2$ as the targeted empirical Bayes estimate of $\sigma_y^2$, our model is as follows.\\\\
For $i=1, \ldots, n$,
$$(y_i, x_i^T)^T \sim \Lambda \eta_i + \varepsilon_i,$$
$$\eta_i \sim N_{\Tilde{k}}(0, I), \,\,\,\,\,\, \varepsilon_i \sim N(0, \Sigma), \,\,\,\,\,\, \Sigma = \text{diag}(\hat{\sigma}_y^2, \sigma_1^2, \ldots, \sigma_p^2),$$
$$\sigma_1^2, \ldots, \sigma_p^2 \sim IG(1,0.3)$$
For $j=1,\ldots, p$ and $l=1, \ldots, \Tilde{k}$,
$$\lambda_{jl} \sim N(0, \xi_{jl}^{-1} \tau_l^{-1}),$$
$$\xi_{jl} \sim IG(1.5, 1.5), \,\,\,\,\,\, \tau_l = \prod_{m=1}^l \delta_l,$$
$$\delta_1 \sim Gamma(2.1,1), \,\,\,\,\,\, \delta_2, \ldots, \delta_{\Tilde{k}} \sim Gamma(3.1,1)$$

In contrast, the competitor we refer to as the ``joint Bayesian factor model'' throughout the paper has the same specification and priors as the above, except that $\sigma_y^2$ is given an $IG(1,0.3)$ prior and is sampled along with the rest of the parameters, rather than being estimated with our empirical Bayes approach.

\section{NHANES predictive MSE for ntrain=50 and ntrain=1600}

\begin{figure}[H]
\centering
\includegraphics[width=0.75\textwidth]{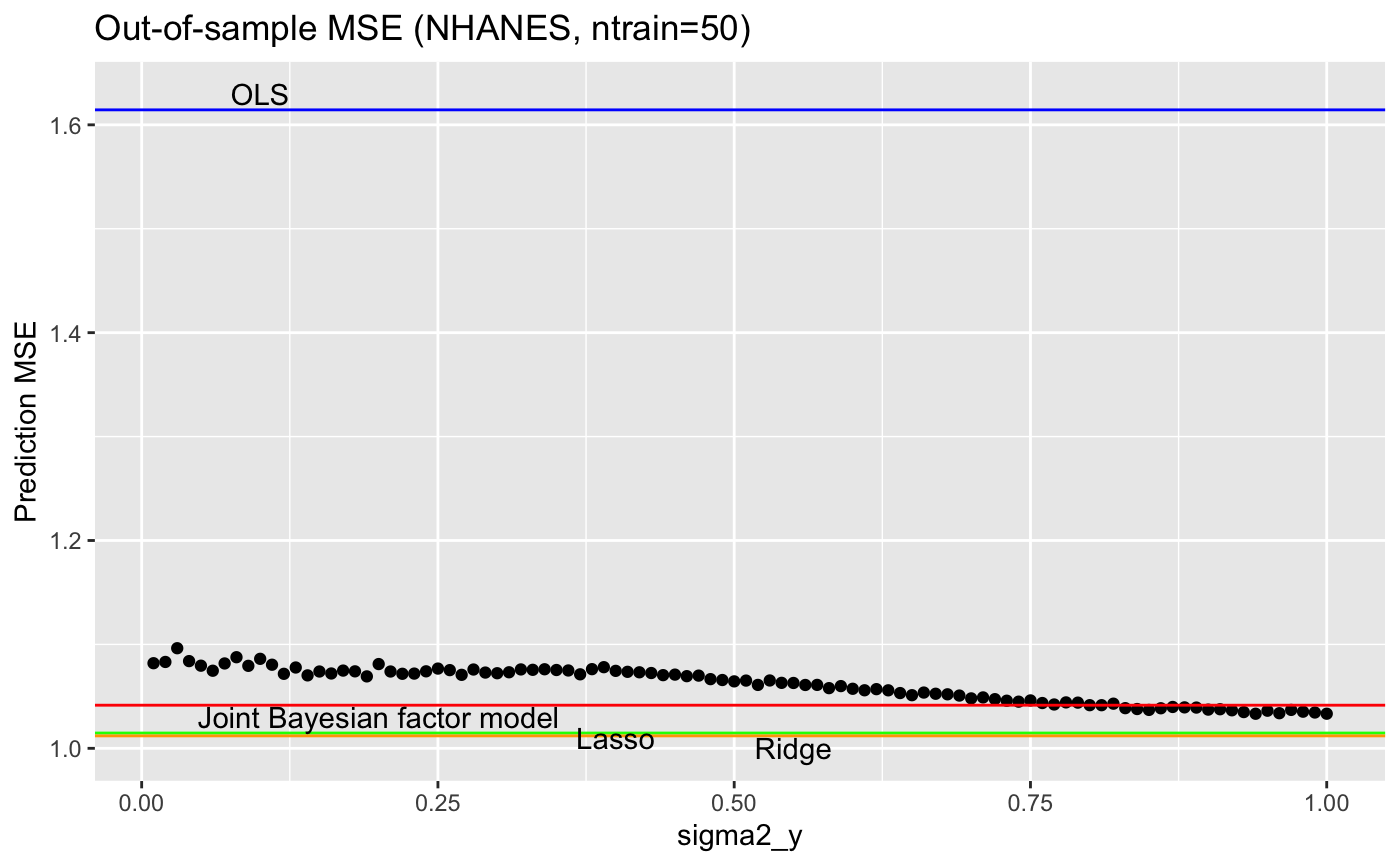}
\caption{\small{Out-of-sample prediction MSE for NHANES data with ntrain=50.}}
\label{fig_nhanes_50}
\end{figure}

\begin{figure}[H]
\centering
\includegraphics[width=0.75\textwidth]{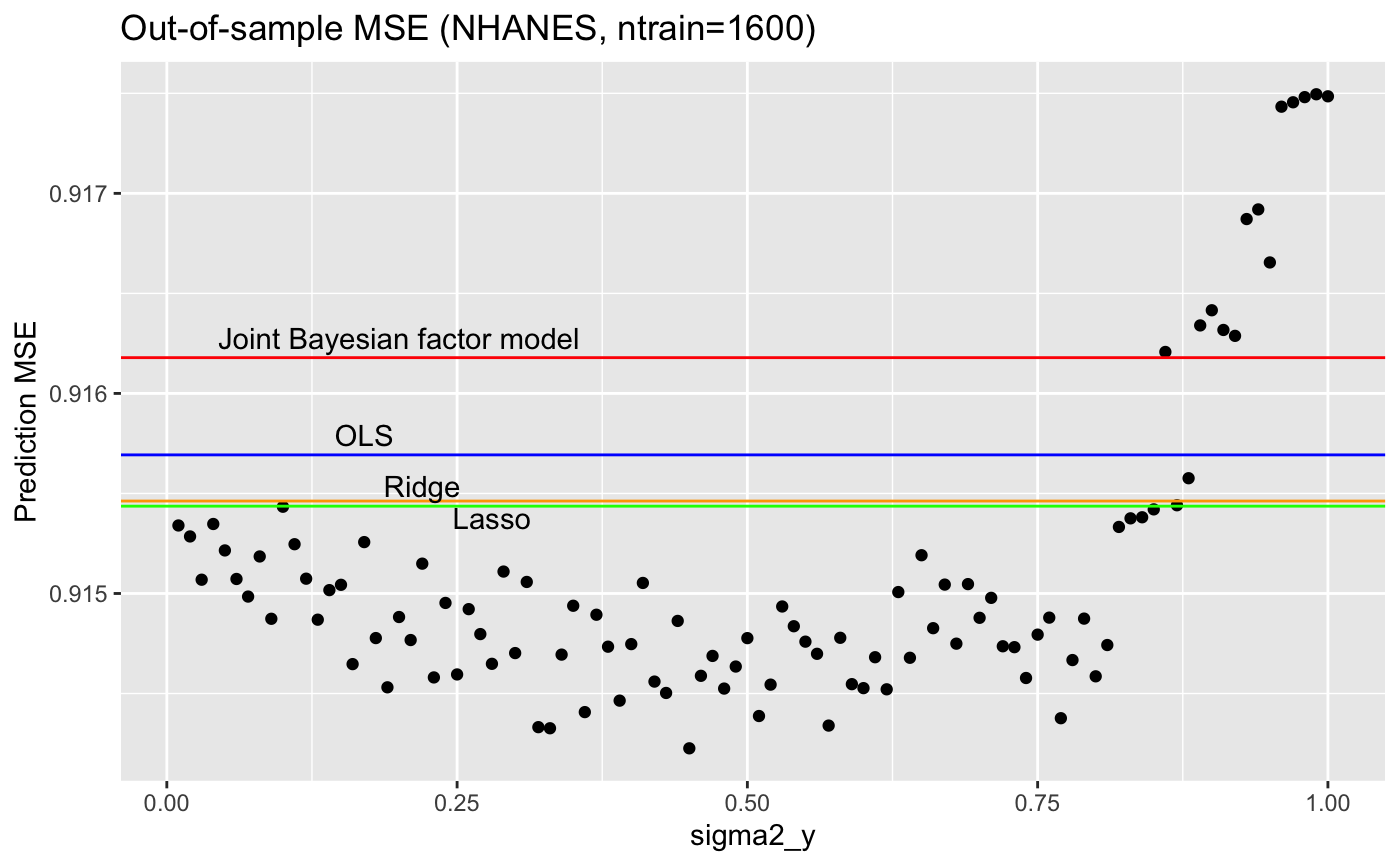}
\caption{\small{Out-of-sample prediction MSE for NHANES data with ntrain=1600.}}
\label{fig_nhanes_1600}
\end{figure}

\newpage

\section{NHANES top 4 columns of $\Lambda$ by outcome loading}
Posterior mean columns of $\Lambda$ learned by TEB-FAR with $\sigma^2_y = 0.5$ (left) and a joint Bayesian factor model with multiplicative gamma process prior on $\Lambda$ (right). In both cases, these are all the aligned columns with $\lambda_y > 0.01$. The loadings for the outcome $y_i$ are in the first row of each matrix, followed by $x_1,\ldots,x_{19}.$ Zeros indicate values for which the $95\%$ posterior credible interval included zero. 
$$
\begin{matrix}
    BMXBMI\\
    URXCNP\\
    URXCOP\\
    URXECP\\
    URXECPT\\
    URXHIBP\\
    URXMBP\\
    URXMC1\\
    URXMCOH\\
    URXMEP\\
    URXMHBP\\
    URXMHH\\
    URXMHHT\\
    URXMHNC\\
    URXMHP\\
    URXMIB\\
    URXMNP\\
    URXMOH\\
    URXMONP\\
    URXMZP
\end{matrix}
\begin{bmatrix} 
0.702 & 0.037 & 0.035 & 0.032 \\ 
0.075 & 0.244 & 0.65 & 0.246 \\ 
0.067 & 0.223 & 0.413 & 0.684 \\ 
0.057 & 0.276 & 0.192 & 0.159 \\ 
-0.045 & 0.055 & 0.142 & 0.131 \\ 
0 & 0.538 & 0.103 & 0.108 \\ 
0.035 & 0.806 & 0.077 & 0.124 \\ 
0.054 & 0.419 & 0.254 & 0.47 \\ 
0.032 & 0.076 & 0.056 & 0.087 \\ 
0.062 & 0.311 & 0.156 & 0.072 \\ 
-0.041 & 0.876 & 0 & 0.096 \\ 
0.044 & 0.284 & 0.113 & 0.14 \\ 
0.062 & 0.162 & 0.063 & 0.156 \\ 
0 & 0.103 & 0.057 & 0.118 \\ 
-0.063 & 0.115 & 0 & 0.197 \\ 
0.057 & 0.47 & 0.111 & 0.135 \\ 
0 & 0 & 0 & 0.82 \\ 
0.052 & 0.301 & 0.115 & 0.131 \\ 
0.064 & 0.258 & 0.185 & 0.777 \\ 
0.118 & 0.524 & 0.164 & 0.134 \\ 
\end{bmatrix}, \,\,\,\,\,\,\,\,\,\,
\begin{matrix}
    BMXBMI\\
    URXCNP\\
    URXCOP\\
    URXECP\\
    URXECPT\\
    URXHIBP\\
    URXMBP\\
    URXMC1\\
    URXMCOH\\
    URXMEP\\
    URXMHBP\\
    URXMHH\\
    URXMHHT\\
    URXMHNC\\
    URXMHP\\
    URXMIB\\
    URXMNP\\
    URXMOH\\
    URXMONP\\
    URXMZP
\end{matrix}
\begin{bmatrix} 
0.347 & 0.031 & 0.032 & 0.032 \\ 
0.154 & 0.236 & 0.64 & 0.253 \\ 
0.145 & 0.214 & 0.373 & 0.69 \\ 
0.124 & 0.268 & 0.165 & 0.16 \\ 
-0.085 & 0.051 & 0.152 & 0.137 \\ 
0 & 0.534 & 0.097 & 0.109 \\ 
0.096 & 0.799 & 0.058 & 0.121 \\ 
0.152 & 0.41 & 0.218 & 0.47 \\ 
0 & 0.072 & 0.051 & 0.089 \\ 
0.115 & 0.304 & 0.141 & 0.071 \\ 
-0.092 & 0.871 & 0.049 & 0.099 \\ 
0.1 & 0.277 & 0.092 & 0.138 \\ 
0.117 & 0.157 & 0 & 0.158 \\ 
0 & 0.097 & 0.052 & 0.12 \\ 
-0.141 & 0.111 & 0 & 0.193 \\ 
0.113 & 0.462 & 0.092 & 0.132 \\ 
0 & 0 & 0 & 0.805 \\ 
0.1 & 0.294 & 0.094 & 0.129 \\ 
0.147 & 0.247 & 0.147 & 0.778 \\ 
0.213 & 0.518 & 0.132 & 0.135 \\ 
\end{bmatrix}
$$

\section{Estimated idiosyncratic variance terms -- NHANES}

\begin{figure}[H]
\centering
\includegraphics[width=0.8\textwidth]{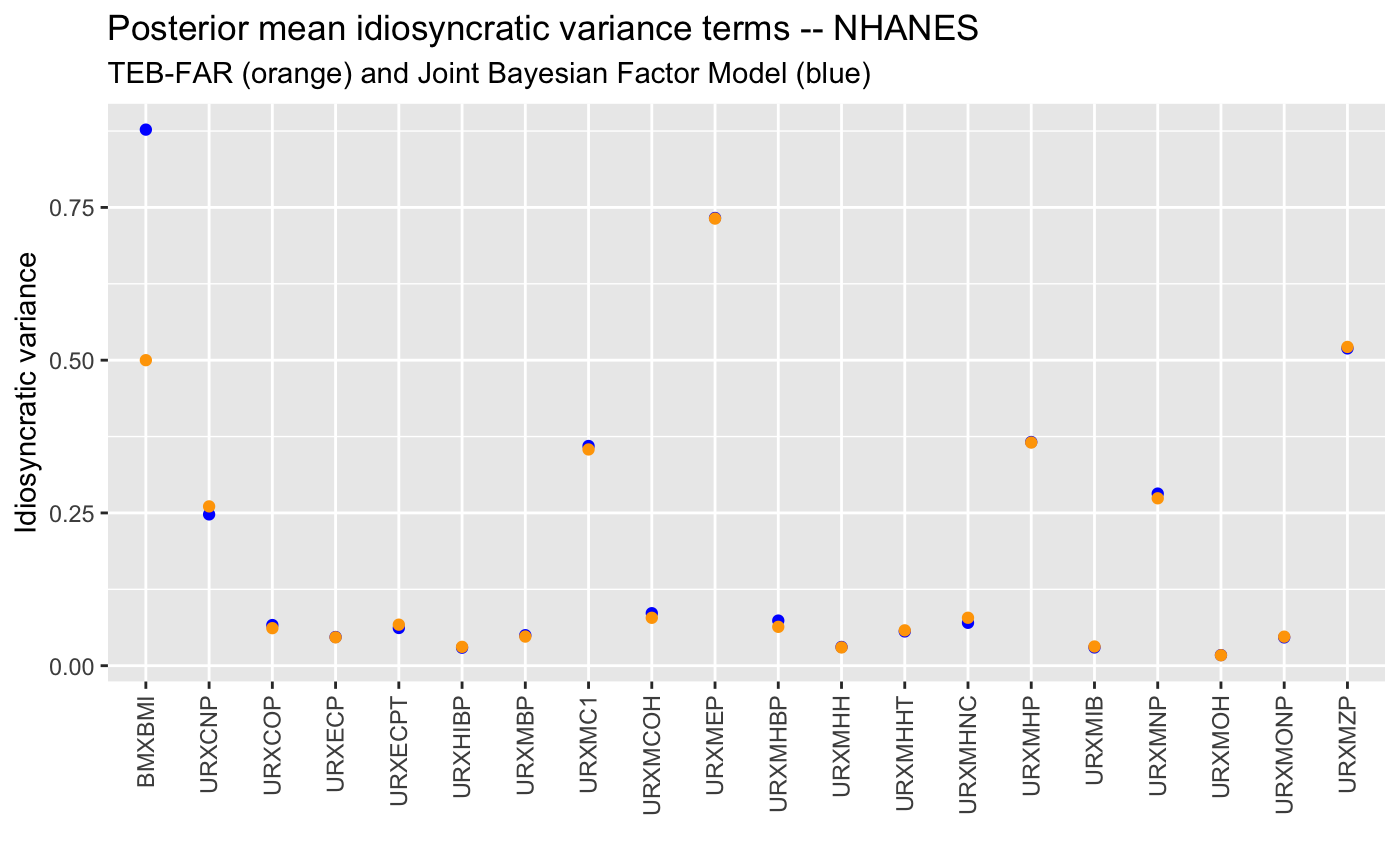}
\caption{\small{Posterior mean idiosyncratic variances estimated by TEB-FAR with $\sigma^2_y = 0.5$ (orange) and a joint Bayesian factor model with multiplicative gamma process prior on $\Lambda$ (blue).}}
\label{fig_idiosyncratic_variances}
\end{figure}

\section{Estimated covariance between BMI and phthalate exposures -- NHANES}

\begin{figure}[H]
\centering
\includegraphics[width=0.8\textwidth]{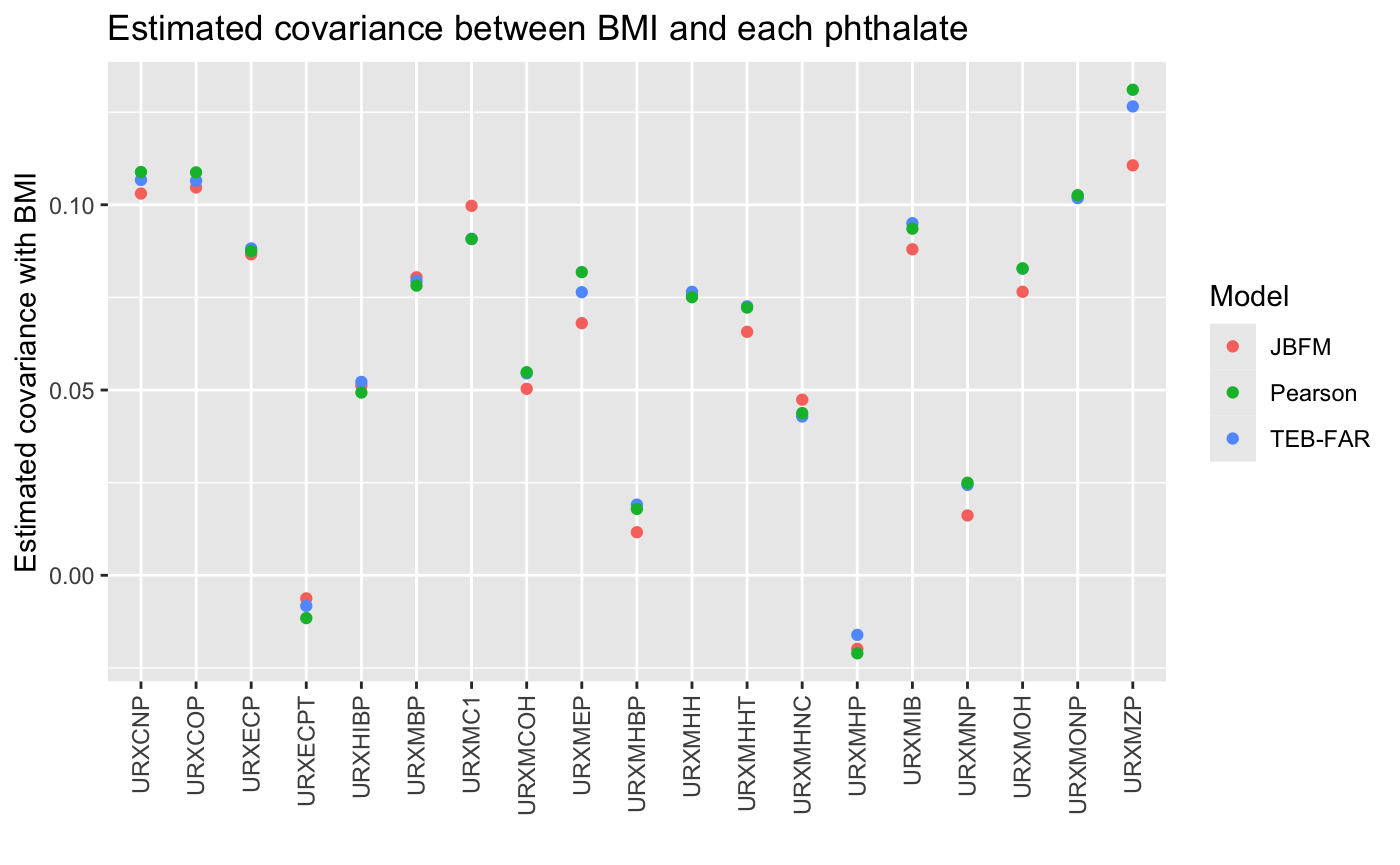}
\caption{\small{Posterior mean covariance terms between BMI and each of 19 phthalates estimated by TEB-FAR with $\sigma^2_y = 0.5$ (green) and a joint Bayesian factor model (JBFM) with multiplicative gamma process prior on $\Lambda$ (blue), compared to the Pearson sample covariances (red). The mean squared difference between JBFM and Pearson is $1.1 \times 10^{-3}$, while the mean squared difference between TEB-FAR and Pearson is $1.1 \times 10^{-4}$, suggesting that by forcing $\sigma_y^2$ to be smaller, we have reduced the amount of regularization applied to learning the joint distribution.}}
\label{fig_cov_BMI_phthalates}
\end{figure}
